\documentclass[12pt]{article}

\usepackage{amsmath}
\usepackage{amssymb}
\usepackage{epsfig}
\usepackage{a4wide}

\newcommand{\NN}{\mathbb N}
\newcommand{\RR}{\mathbb R}

\begin{document}

\setlength{\parindent}{15pt}

\vspace*{3em}
\centerline{\Large \bf Signal analysis of impulse response functions }
\vspace{0.7em}
\centerline{\Large \bf  in MR- and CT-measurements of cerebral blood flow}

\vspace{10mm}

\centerline{\sc Evelyn Rost$^{\rm a,c}$, Ralf Geske$^{\rm b}$ and 
Michael Baake$^{\rm a}$}
\vspace{10mm} 

{\footnotesize \noindent
\begin{itemize}
  \item[(a)] Fakult\"at f\"ur Mathematik, Univ.\ Bielefeld, Box 10 01 31,
      33501 Bielefeld, Germany
  \item[(b)] Institut f\"ur klinische Physik und Biomedizintechnik, 
      Dietrich Bonhoeffer Klinikum \\Neubrandenburg, Allendestr. 30, 
      17036 Neubrandenburg, Germany
  \item[(c)] Institut f\"ur Mathematik und Informatik, Univ.\ Greifswald,
      Jahnstr.\ 15 a,\\ 17487 Greifswald, Germany
\end{itemize}
}


\vspace{10mm}
\begin{abstract}
\noindent
The impulse response function (IRF) of a localized bolus in cerebral
blood flow codes important information on the tissue type. It is
indirectly accessible both from MR- and CT-imaging methods, at least
in principle. In practice, however, noise and limited signal
resolution render standard deconvolution techniques almost useless.
Parametric signal descriptions look more promising, and it is the aim
of this contribution to develop some improvements along this line.
\end{abstract}

\vspace{10mm}
\noindent
Key Words: blood flow, tissue determination, parametric fits, \newline
\hphantom{Key Words:} Fourier transform, MR- and CT-imaging



\section{Introduction}

The detailed knowledge of blood circulation in the brain is beyond
doubt an important diagnostic tool. Among other aspects, it admits the
detection and a partial characterization of cerebro vascular and
tumorous diseases. Apart from the relative cerebral blood volume, the
mean transit time is a basic diagnostic parameter. For its
determination, the so-called impulse response function (sometimes also
called impulse residue function) \cite{lA83, rW00} of the tissue is
needed, which, unfortunately, is not accessible to direct measurement.
Previous contributions consistently showed various deficiencies in
regard to its determination, especially if diffusive tissue areas were
involved.

The aim of this contribution is to analyze such deficiencies from a
more mathematical point of view and to point out a method that is
practically feasible. Both MR and CT perfusion measurements are
included here.

First, we notice that the determination of the impulse response
function of tissue by a direct deconvolution from the arterial and
tissue signal does not seem to be possible in a reliable way. A
solution is then suggested on the basis of a parametric fit. In the
case of a disturbed blood brain barrier, it is shown that an
estimation of the mean transit time is possible also with CT perfusion
measurements, although a CT measurement is short in comparison to an
MR measurement since CT measurement is time limited due to radiation
exposure.

\medskip The paper is organized as follows. First, we briefly
summarize the deconvolution methods used, and their common
deficiencies in our practical situation. We continue with a more
detailed discussion of parametric fits. Here, we discard the so-called
double gamma fit, because it can be shown to lead to inconsistent
results. Nevertheless, a closely related pair of parametric functions
works rather well and is suggested as an alternative.

We describe first steps towards the parametric treatment of data from
defective tissue, and close with the suggestion of a quantity to
distinguish intact from disturbed blood brain barrier. This is the
standard deviation of the IRF and can replace the mean transit time.
It is directly accessible even without a parametric fit, both for MR
and CT data.

\section{Shortcomings of convolution techniques}

Three numerical deconvolution methods were analyzed in \cite{Rost}.
They are deconvolution by
\begin{enumerate}
\item Fourier transform
\item the method of moments
\item functional calculus.
\end{enumerate}
These methods were first tested by realistic, but simulated data
without noise. They provide acceptable and comparable results if the
sampling rate is sufficiently high (which roughly means some five to
ten sampling points per relevant interval of time). Let us explain the
deconvolution approach in some more detail.

\medskip 1. The deconvolution by Fourier transform uses the
convolution theorem \cite[p.~110]{rB86}, so that one gets
\begin{equation}\label{entfalt}
g=\biggl{(}\frac{(f*g)\ \hat{}\ }{f\ \hat{}}\biggr{)}^{\displaystyle\check{}},
\end{equation}
where $(.)\ \hat{}\ $ denotes the Fourier transform and $(.)\
\check{}\ $ the corresponding inverse transform. One principle problem
is the following. Whenever the Fourier transform of $f$ and $g$ are
equal in a certain interval including zero, and the Fourier transform
of $f$ is zero outside this interval, then $(f*f)\ \hat{}=(f*g)\
\hat{}\ $. In general, it is thus not correct that $f*g_1=f*g_2$
implies $g_1=g_2$. This means that the deconvolution is not well
defined in general, compare \cite[Ch.~7]{mK87} for a general
discussion. Fortunately, in our present context, this aspect does not
play a major role. Another numerical problem, which is more important
in our case, can emerge from (approximate) zeros of the denominator of
\eqref{entfalt}. It is obvious that the zeros of $f\ \hat{}\ $ form a
subset of those of $(f*g)\ \hat{}\ $, and that \eqref{entfalt} is
always well defined from the theoretical point of view. However, since
the Fourier transforms have to be determined numerically from discrete
data, it can occur that the approximation of $f\ \hat{}\ $ features
zeros where that of $(f*g)\ \hat{}\ $ does not. This makes expression
\eqref{entfalt} undefined or leads to undesirable fluctuations.
Finally, the numerical data for $f$ and for $f*g$ show similar noise
level. In theory, $f*g$ should be the smoother signal though, and a
violation can (and actually does) lead to numerical sensitivity of the
(discrete) inverse transform.\medskip

2. If the function $g$ (or a good approximation of it) is known in
some parametric form, the unknown parameters can be determined by the
method of moments. To this end, we need the following relation
\cite{Rost}: If $f$ and $g$ are non-negative, suitably integrable,
normalized functions, then
\begin{equation}\label{nmom}
  \bigl{<}X^n\bigr{>}_{f*g}=\sum\limits_{k=0}^{n}\binom{n}{k}\bigl{<}
   X^{n-k}\bigr{>}_g\bigl{<}X^k\bigr{>}_f\ ,
\end{equation}
where $\bigl{<}X^n\bigr{>}_h:=\int x^nh(x)dx$ is the $n$-th moment of
$h$. In particular, we assume that all these moments exist, which is
very reasonable in our context.

Assume that it is possible to describe $g$ by $m$ parameters. In this
case, all moments up to order $(m-1)$ of $f$ and $f*g$ are needed to
create a complete set of equations by means of \eqref{nmom}. In our
case, the functions $f$ and $f*g$ are non-negative, but not normalized
(this can be mended by dividing by their respective integrals). Since,
however, the moments have to be calculated numerically, they will show
decreasing accuracy with increasing $m$.  Consequently, this approach
is only useful for situations with a small number of parameters.\medskip

3. Deconvolution by the functional calculus is a less well known, but
rather interesting method. It provides a kind of inverse formula to
determine the function $g$ if $f$ and $f*g$ are given. This does not
work for any arbitrary function $f$ but $f$ has to be a special type
of function. In principle, the method applies to our situation, which
had been overlooked in previous attempts. The following example shall
illustrate this method. If
\begin{equation}\label{fdef}
f(t)=\begin{cases}
te^{-t}, \quad t\geq 0,\\
0, \qquad \text{otherwise},
\end{cases}
\end{equation}
the equation $(f*g)(t)=\int_{\RR}f(t-x)g(x)dx$ is inverted by
\begin{equation}
g(t)=(1+D)^2(f*g)(t),
\end{equation}
where $D=\frac{d}{dt}$ is the derivation operator. This means that $g$
can be calculated by the first and the second derivative of $f*g$, if
$f$ is according to \eqref{fdef}. Further examples, and a systematic
theory, can be found in \cite{iH55}.  Unfortunately, derivatives play
a major role, and, with increasing complexity of the convolution
kernel, the derivatives required are of higher and higher order. Since
the actual data are discrete values, the derivatives have to be
calculated numerically. Hence, this method is suitable only to a
limited extent due to numerical sensitivity.\medskip

\begin{figure*}
\centerline{\epsfig{file=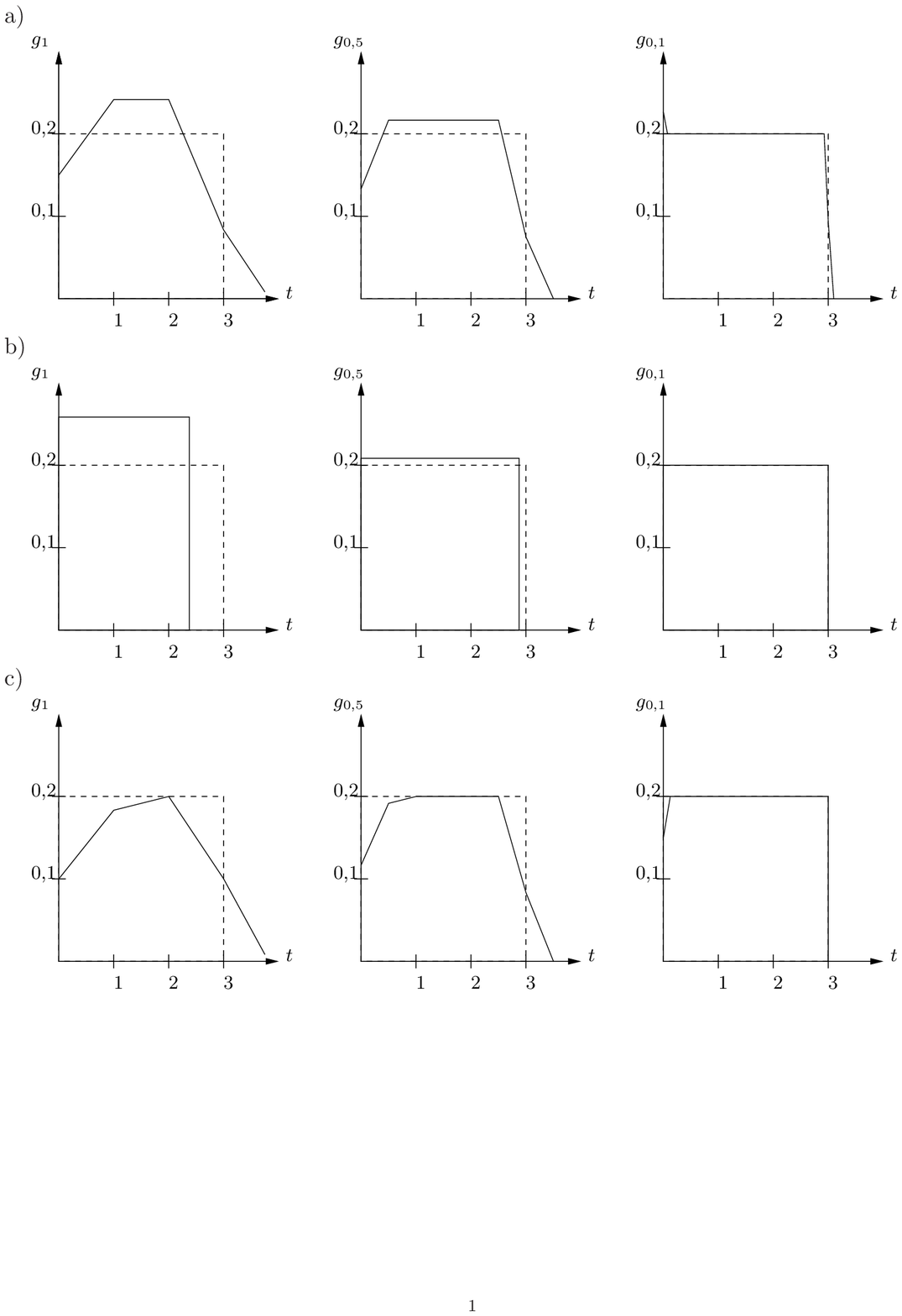}}
\caption{\small \sl \label{ergeb}Results of deconvolution by a) Fourier transform
  b) the method of moments and c) the functional calculus in
  dependence on the sampling rate. The latter is one point (left), two
  points (center) and ten points (right) per relevant interval of time
  (which is 1 in this case). For better illustration, the discrete
  sample points are successively connected by straight lines without
  additional interpolation. Note, that the function $g$ is not
  normalized.}
\end{figure*}

As a common problem to all three methods, the sampling rate of about
one point per second of an MR and a CT scanner is too coarse. This is
demonstrated by the following simple example. The function
$f(t)=te^{-t}$, $t\geq0$, was convolved with a rectangular signal $g$
of width 3 and height 0.2 (hence not normalized) to obtain $f*g$.
Then, the particular functions were sampled at different rates. The
results of deconvolution by the three methods are shown in Figure
\ref{ergeb}, whereas the dashed rectangle is the original function $g$
for comparison.

In comparison to this illustrative example, our real world case
corresponds to the leftmost column in Figure \ref{ergeb}, i.e., to a
clearly insufficient resolution. Note that the same qualitative
picture would emerge for other signal types as well, not just for
rectangular $g$.\medskip

Additionally, the real data are superimposed by noise. From a careful
inspection of our sample data, we estimate that noise and measurement
errors together result in fluctuations of five to ten percent on
average. It is well known that numerical deconvolution processes can
be very sensitive to noise. Therefore, it is preferable to describe
the measured data by functions, e.g., by means of a parametric ansatz
followed by a least squares fit. This requires an adequate approach.

\section{Inadequacy of the double gamma fit}

Fitting the data by gamma densities (in the literature mostly called
gamma variate functions \cite{tB97, dS98}) seems to be an adequate
approach. Especially the measured arterial signals can be described
this way. But sometimes both arterial and tissue signal were displayed
by gamma densities \cite{jT, rW00}. In this section, we want to show
that such a double gamma fit is not reasonable.

For simplicity, we formulate the situation in terms of probability
theory. The actual signals are described by non-negative functions
that need not be normalized. However, the normalization constants play
no role here (one can simply divide by them to get normalized
signals), so that we can profit from the standard tools of probability
theory.\medskip

If $f$ is an integrable non-negative function,
\begin{equation}
\phi(t)=\int_{\RR}e^{itx}f(x)dx
\end{equation}
is the (inverse) Fourier transform of $f$. If $f$ is a probability
density, $\phi$ is called its {\em characteristic function}, see
\cite{aS96} for details.  In the case of $f=f_{\lambda, \alpha}$ being
a (normalized) gamma density, i.e.,
\begin{equation}\label{gamma}
f_{\lambda, \alpha}(x)=\begin{cases}
0, \qquad \qquad \qquad \ x<0,\\
\frac{\lambda^\alpha}{\Gamma(\alpha)}x^{\alpha-1}e^{-\lambda x}, \quad \ x\geq0,
\end{cases}
\end{equation}
with $\lambda>0$, $\alpha>1$ and
$\Gamma(\alpha)=\int\limits_{0}^{\infty}x^{\alpha-1}e^{-x}dx$, the
characteristic function is \cite[p.~343]{aS96}
\begin{equation}
\phi_{\lambda, \alpha}(t)=\biggl{(}\frac{\lambda}{\lambda-it}
\biggr{)}^\alpha, \quad \text{with}\ t\in{\RR},
\end{equation}
which can be calculated with elementary methods from complex analysis.

Starting from two gamma densities, we consider the convolution
$f_{\lambda, \alpha}*g=f_{\gamma, \nu}$. (In doing so, $f_{\lambda,
  \alpha}$ corresponds to the measured arterial signal and $f_{\gamma,
  \nu}$ to the tissue signal, whereas $g$ is the IRF wanted, up to a
multiplicative factor). By means of the convolution theorem and the
characteristic function of the gamma density, one gets the
characteristic function of $g$ as
\begin{equation}
\phi_g(t)=\frac{\nu^\gamma(\lambda-it)^\alpha}{(\nu-it)^\gamma\lambda^\alpha}=:h(t).
\end{equation}
The central question now is whether $h$ is the characteristic function
of a probability density. An answer is given by the Bochner-Khintchine
theorem which states that a continuous function $\varphi(t)$,
$t\in{\RR}$, with $\varphi(0)=1$ is the characteristic function of a
probability measure if and only if it is positive definite, see the
appendix for details.

The violation of some elementary properties of positive definite
functions in this case already implies that $h$ is not positive
definite in general. Fitting the data by gamma densities provides
parameters in a region where property (3) of the appendix is violated,
see \cite{Rost} for details. Consequently, $h$ is not a characteristic
function and therefore $g$ is not a probability density. This, in
turn, means that $g$ takes negative values which cannot be interpreted
in our context. Consequently, the sometimes used double gamma fit is
not consistent in this context and should be avoided.

\section{An alternative approach}\label{approach}

Let us look for an alternative. It is reasonable (and also rather
plausible) to assume the IRF itself to be gamma distributed. In this
case, we can calculate the shape of the tissue signal if the arterial
signal is still described by a gamma density. We have to determine
$f*g$ provided that $f=f_{\lambda, \alpha}$ and $g=g_{\mu, \beta}$
according to \eqref{gamma}. Once again, by means of the convolution
theorem and the characteristic function of the gamma density, one now
finds
\begin{equation}
  \phi_{f*g}(t)
  =\biggl{(}\frac{\lambda}{\lambda-it}\biggr{)}^\alpha
  \biggl{(}\frac{\mu}{\mu-it}\biggr{)}^\beta
  =\lambda^\alpha\mu^\beta(\lambda-it)^{-\alpha}(\mu-it)^{-\beta}\ ,
\end{equation}
with $\phi_{f*g}(0)=1$. The corresponding inverse transform reads \cite[Ch.~3.2]{aE54}
\begin{equation}\label{hypfit}
\begin{split}
(f*g)(x)
&=\lambda^\alpha\mu^\beta\ \frac{e^{-\lambda x}x^{\alpha+\beta-1}}
{\Gamma(\alpha+\beta)}{}_1F_1[\beta;\alpha+\beta;(\lambda-\mu)x]\\
&=\lambda^\alpha\mu^\beta\ \frac{e^{-\mu x}x^{\alpha+\beta-1}}
{\Gamma(\alpha+\beta)}{}_1F_1[\alpha;\alpha+\beta;(\mu-\lambda)x]
\end{split}
\end{equation}
for $x\geq0$ and $(f*g)(x)=0$ otherwise, where $_1F_1$ is the
confluent hypergeometric function \cite[Ch.~13]{mA72} with
\begin{equation*}
{}_1F_1(a;b;z)
:=\sum\limits_{k=0}^\infty\frac{(a)_k}{(b)_k}\frac{z^k}{k!}
\ =1+\frac{az}{b}+\frac{a(a+1)}{b(b+1)}\frac{z^2}{2!}+\ldots\ ,
\end{equation*}
whereas $(c)_k:=c(c+1)\cdot\ldots\cdot(c+k-1)$ is the Pochhammer
symbol. The normalization of the inverse transform is chosen such that
$(f*g)(x)$ is indeed a normalized probability density, i.e.,
$(f*g)(x)\geq0$ together with $\int_{{\RR}_+}(f*g)(x)dx=1$.

The two formulations of \eqref{hypfit} are equal by Kummer's
functional equation for ${}_1F_1$, see \cite[Eq.~13.1.27]{mA72}. For
$\lambda>\mu$ (resp. $\lambda<\mu$) the first (resp. the second)
version immediately shows the non-negativity of $(f*g)(x)$. For
$\lambda=\mu$, $f*g$ is the density of a gamma distribution, in line
with the convolution semigroup property $(f_{\lambda,
  \alpha}*f_{\lambda, \beta})(x)=f_{\lambda, \alpha+\beta}(x)$ for
gamma densities with the same parameter $\lambda$, compare
\cite{rD91}.

\begin{figure}[!ht]
\centerline{\epsfig{file=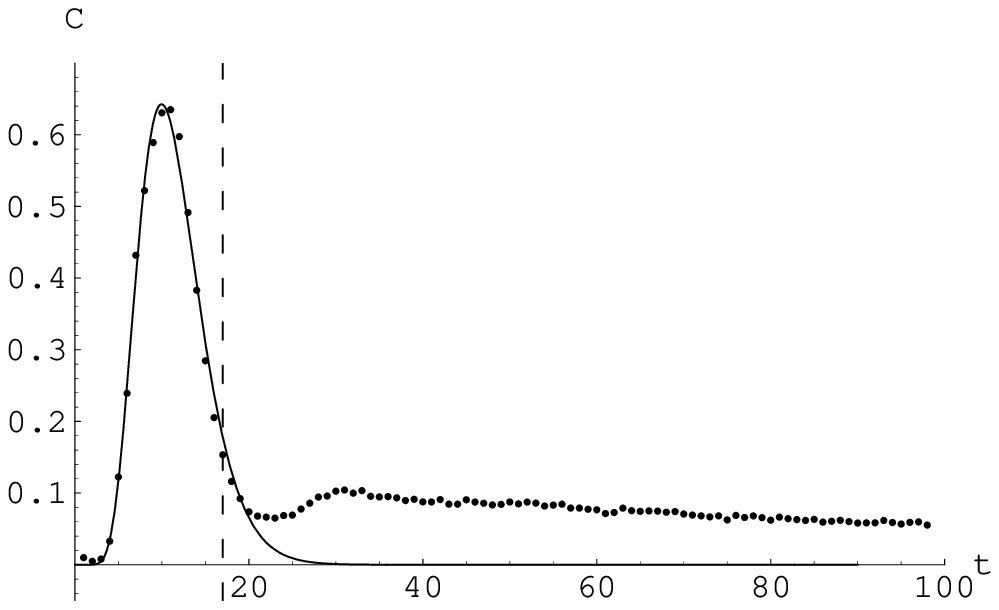,width=70mm},\epsfig{file=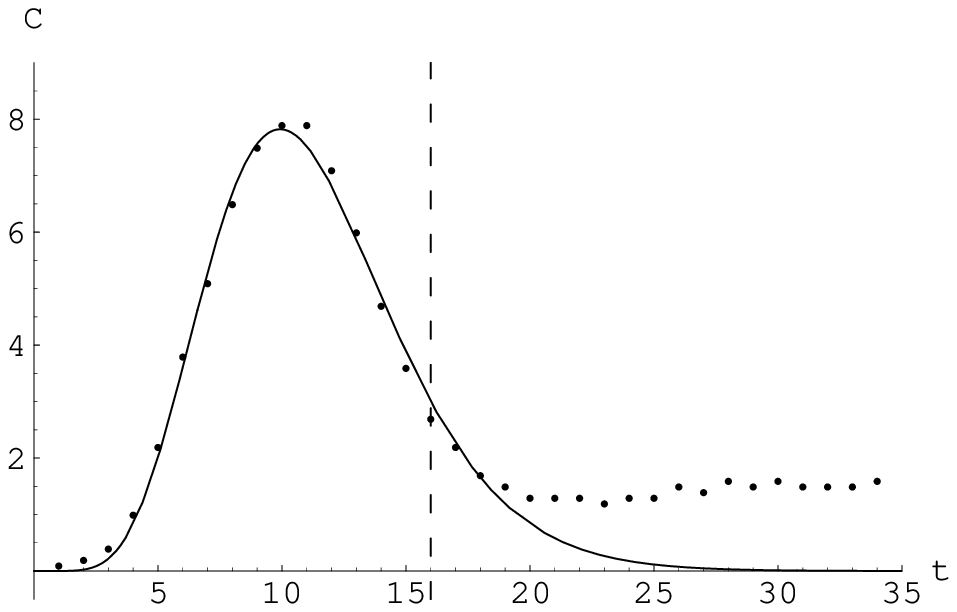, width=70mm}}
\caption{\small \sl \label{hypfitbild} Parametric fit of the tissue signal
  measured by an MR scanner (top) and a CT scanner (bottom), with the
  function from Eq.~\eqref{hypfit}, up to an additional normalization
  factor. All points to the left of the dashed vertical line were used
  for the fit.}
\end{figure}

The question is now whether the tissue signal can be described by the
expression in \eqref{hypfit}. An example (compare with Figure
\ref{hypfitbild}) shows that the fit is rather convincing, so that
this ansatz is phenomenologically sound. Of course, it remains an
interesting open question at this point whether further corroboration,
e.g., by microscopic modelling, is possible.\medskip

The function in Eq.~\eqref{hypfit} fits the tissue signal at least as
well as a gamma density would do. But in contrast to the latter, we no
longer encounter problems with negative values of the IRF.

Note, however, that this approach can only be used for tissue with an
intact blood brain barrier because the fit separates the recirculation
from the real signal. Tissue with a defective blood brain barrier,
such as a brain tumor, has to be considered separately because
recirculation is superimposed by diffusion.

\subsection{Tissue with defective blood brain barrier}

The measured signal of a tissue with a permeable blood brain barrier
(e.g., brain tumor) consists of a diffusive and a perfusive part
\cite{aB98, aC00}. The idea here is to split up these parts and
deconvolve them separately. This way, additional features of the
different contributions can be detected and displayed more clearly.

After injection of the contrast medium, most of these molecules remain
in the intravascular space during their transit through the vascular
network. However, a small portion diffuses into the extravascular
space and then back into the intravascular space due to the defective
blood brain barrier. Because diffusion is much slower than perfusion,
contrast medium molecules still diffuse into the extravascular space
during recirculation \cite{aB98, aC00, dS98}.

\begin{figure}[!ht]
\centerline{\epsfig{file=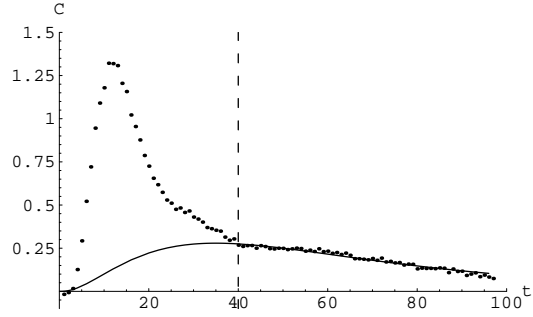,width=70mm}}
\caption{\small \sl \label{difbild}The function $f(t)=at^be^{-ct^d}$ as a
  parametric fit for the diffusive part. All points to the right of
  the dashed vertical line were used to determine the associated
  parameters.}
\end{figure}

To our knowledge, there is no theoretical derivation known to describe
the diffusive part in detail, wherefore a phenomenological approach is
necessary and appropriate. The function
\begin{equation}\label{diffit}
f(t)=\begin{cases}
0,  \qquad \qquad  \: \: t<0,\\
at^{b}e^{-c t^d}, \quad \ \ t\geq0,
                     \end{cases}
\end{equation}
with $a>0$, $b\geq1$, $c>0$ and $d>0$ unknown parameters, seems to be
applicable to capture the diffusion part parametrically, see Figure
\ref{difbild}.

\begin{figure}[!ht]
\centerline{\epsfig{file=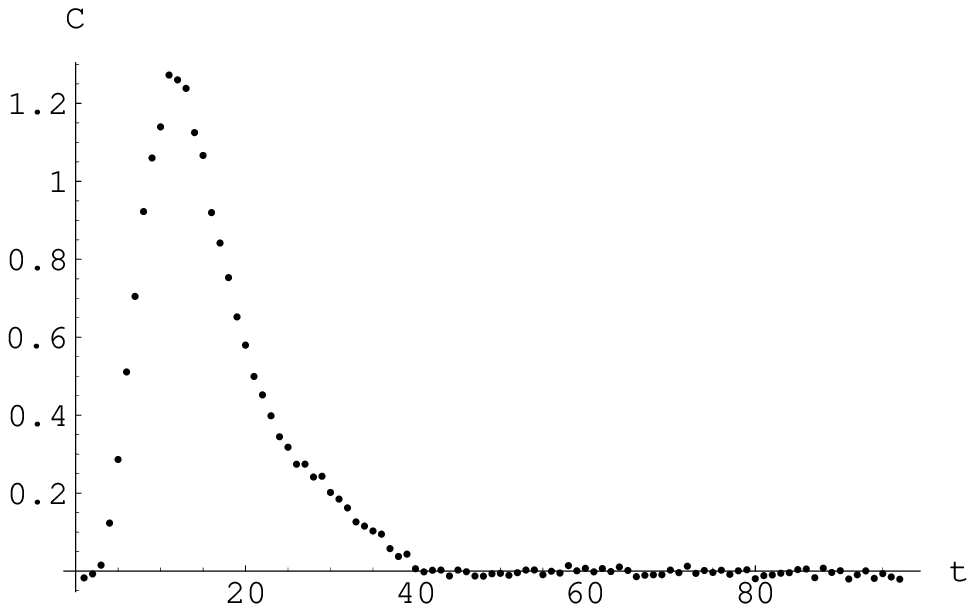,width=70mm},\epsfig{file=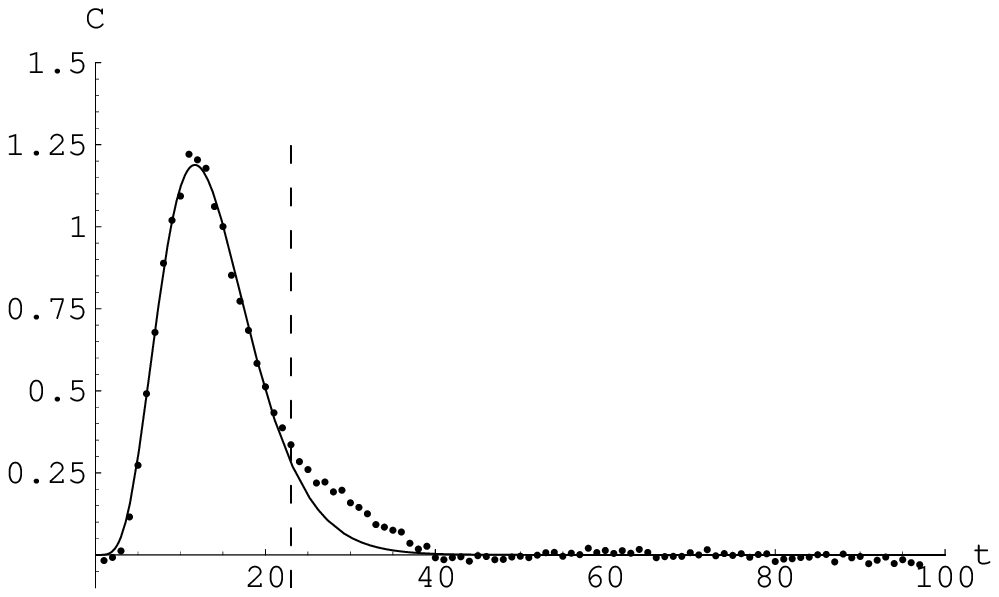, width=70mm}}
\caption{\small \sl \label{perf}Perfusion after subtraction of the diffusive part
  (top) and Eq.~\eqref{hypfit} as an adequate fit of the perfusion
  (bottom). All points to the left of the dashed line were used in
  this fit. The few points to the right not fitted by \eqref{hypfit}
  are the remainder of recirculation.}
\end{figure}

The perfusive part can be obtained approximately by subtraction of the
diffusion part from the original signal (compare with Figure
\ref{perf} top). Figure \ref{perf} bottom shows that \eqref{hypfit}
leads to a fit of acceptable quality. This means that the method
introduced above can also be applied to determine the IRF of the
perfusion. In this case, the IRF is a gamma density with the
associated parameters from the fit.

\begin{figure}[!ht]
\centerline{\epsfig{file=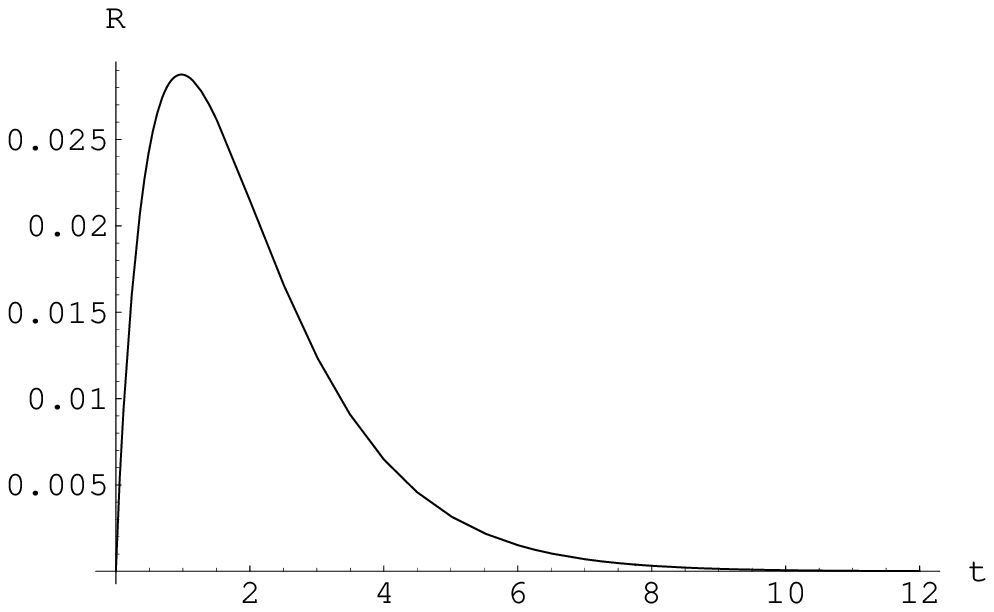,width=70mm},\epsfig{file=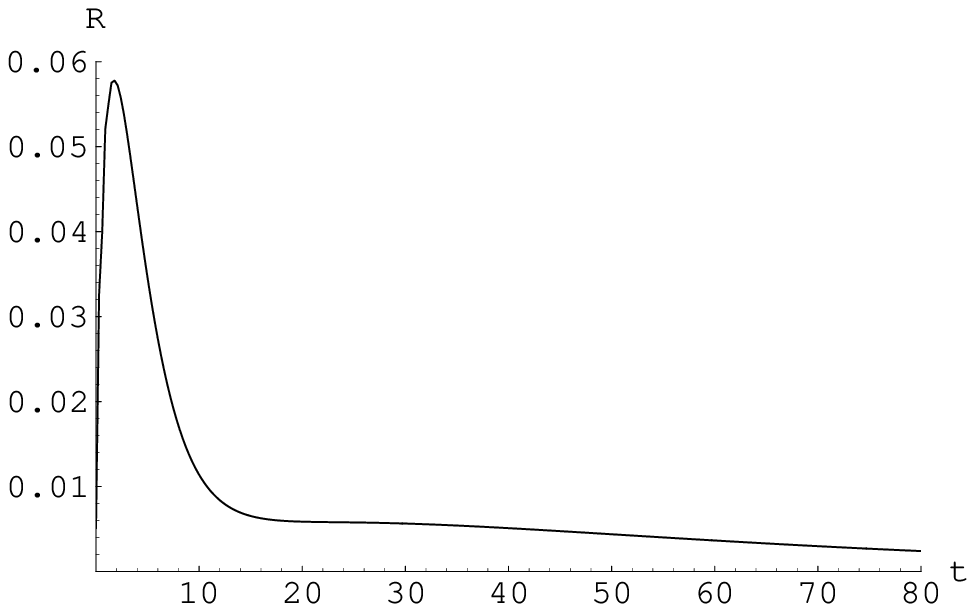, width=70mm}}
\caption{\small \sl \label{irf}The IRF for tissue with an intact blood brain
  barrier (top) and a defective blood brain barrier (bottom). Note the
  different scale on the time axis for comparison.}
\end{figure}

It remains to consider the diffusive part. We did not find an explicit
and practically useful expression for the IRF, so that we employed a
numerical deconvolution technique (e.g., by Fourier transform). This,
in contrast to above, does not pose a problem because both signals
required are now available in parametrized form, and because the
non-uniqueness problem does not show up. Now, one can put the IRFs
together to obtain the IRF of tissue with a permeable blood brain
barrier. Figure \ref{irf} shows the IRFs for the different tissue
types. On the top, the IRF for tissue with an intact blood brain
barrier is shown, and the IRF for tissue with a defective blood brain
barrier on the bottom.

\subsection{Diagnostic parameters}

So far, the {\em area to height relation}\/ has been used to determine
the mean transit time ($MTT$) \cite{aC99, rW00}. There, the area under
the curve of the IRF is divided by the maximum of the IRF, i.e.,
\begin{equation}\label{mtt}
MTT=\frac{\int R(t)dt}{R_{max}},
\end{equation}
if the IRF is denoted by $R$. However, this approach is only
applicable for tissue with an intact blood brain barrier under the
simplifying (and somewhat unrealistic) assumption that the IRF is a
pulse of rectangular shape. By application of this formula to tissue
with a permeable blood brain barrier, a rather unrealistic and
unreliable estimate of the mean transit time is obtained. For
characterizing and possibly detecting a brain tumor, a more reliable
quantity is required.

\medskip The standard deviation of the IRF appears to be an adequate
measure for the mean transit time. To expand on this suggestion, let
$f$ be a non-negative signal with a compact support. If
$N=\int_{\RR}f(t)dt$ is the normalization constant,
\begin{equation}
m_f:=\frac{1}{N}\int_{\RR}tf(t)dt
\end{equation}
is the signal mean and
\begin{equation}\label{sigma}
\sigma_f
=\sqrt{\frac{1}{N}\int_{\RR}(t-m_f)^2f(t)\, dt}\,
=\sqrt{\frac{1}{N}\int_{\RR}t^2f(t)dt-m_f^2}
\end{equation}
is its standard deviation.

The standard deviation features the correct dimension, namely time,
and does not change if the entire signal is shifted along the time
axis. More importantly, its definition is independent of the shape of
the IRF in the sense that it does not require any specific assumptions
on the signal form. Consequently, it is applicable to the different
tissue types.

\medskip In the case of the IRF being a non-normalized gamma density,
there is an asymptotically linear relation between the mean transit
time, computed by the area to height formula, and the standard
deviation of the IRF. Thus, it is no big deal to replace the
traditional formula by the standard deviation, because one calculates
nothing much different in this case.

\begin{figure}[!ht]
\centerline{\epsfig{file=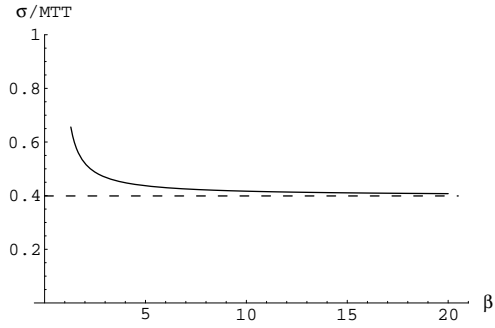,width=70mm}}
\caption{\small \sl \label{mttzustdabw}Theoretical relation between $\sigma$ and
  $MTT$ in the case of a gamma density, in dependence on the parameter
  $\beta$. The dashed line runs at height
  $\frac{1}{\sqrt{2\pi}}\simeq0.39$.}
\end{figure}

The theoretical relation can be derived as follows. Let the IRF again
be denoted by $R$ and assume that it is a non-normalized gamma density
with parameters $b$, $\beta$ and $\mu$, i.e.,
\begin{eqnarray}
R(t)=\begin{cases}
b\frac{\mu^\beta}{\Gamma(\beta)}t^{\beta-1}e^{-\mu t}, \quad t\geq0\\
0, \qquad \qquad \qquad\text{otherwise}.
\end{cases}
\end{eqnarray}
The standard deviation reads $\sigma_\Gamma=\frac{\sqrt{\beta}}{\mu}$,
the area under the curve is $\int_{\RR}R(t)dt=b$ and it is easily
checked that the maximum value is
$R_{max}=R\bigl{(}\frac{\beta-1}{\mu}\bigr{)}=b\frac{\mu^\beta}
{\Gamma(\beta)}\bigl{(}\frac{\beta-1}{\mu}\bigr{)}^{\beta-1}e^{1-\beta}$.
Calculating the ratio of $\sigma_\Gamma$ and $MTT_\Gamma$ from
\eqref{mtt} leads to
\begin{equation}
  \frac{\sigma_\Gamma}{MTT_\Gamma}=\frac{\sqrt{\beta}}{\Gamma(\beta)}
  \frac{(\beta-1)^{\beta-1}}{e^{\beta-1}}.
\end{equation}
This expression depends on parameter $\beta$ only. Figure
\ref{mttzustdabw} shows this function in dependence on $\beta$. The
dashed horizontal line indicates the asymptotic value of the function
as $\beta\rightarrow\infty$.

\medskip
Stirling's formula \cite[Eq.~6.1.37]{mA72} states that
\begin{equation}
\Gamma(x)=e^{-x}\,x^{x-\frac{1}{2}}\sqrt{2\pi}\ (1+\mathcal O(x^{-1}))
\end{equation}
as $x\rightarrow\infty$.
With this expression we get
\begin{equation}\label{propzus}
  \frac{\sigma_\Gamma}{MTT_\Gamma}
=\frac{1}{\sqrt{2\pi}}\
  \frac{\beta}{\beta-1}+{\mathcal O}(\beta^{-1})
\quad \xrightarrow{\beta\to\infty}\quad \frac{1}{\sqrt{2\pi}}\simeq0.39\ .
\end{equation}

Using the standard deviation of the IRF as a measure for the mean
transit time, one gets significantly different values in dependence on
the tissue type. Tissue with a permeable blood brain barrier (e.g.,
brain tumor) possesses a greater mean transit time than normal tissue
(compare with figure \ref{irf} again).\medskip

To turn this into a practical method, the following ansatz looks
promising. It is possible to determine the standard deviation of the
IRF directly from the given (noisy) data without making an explicit
deconvolution. Equation \eqref{nmom} implies
$\sigma_{f*g}^2=\sigma_f^2+\sigma_g^2$, if $f$ and $g$ are probability
densities. Consequently, the standard deviation of $g$ can be
calculated as
\begin{equation}\label{sigdis}
\sigma_g=\sqrt{\sigma_{f*g}^2-\sigma_f^2}\ .
\end{equation}

In our case, we have to normalize the signals prior to applying
\eqref{sigdis}. This can be done on the basis of the discrete data
directly. Our signals are available as a finite time series $\{h_\ell\
\vert\ 1\leq\ell\leq n_{max}\}$, where we assume, for simplicity, that
the integral index $\ell$ is the time (this being literally true for
our MR test data). The normalization leads to
\begin{equation}
p_\ell=\frac{h_\ell}{\sum_j h_j}\ ,
\end{equation}
where $h_k$ are the non-normalized values of the respective signal and
$p_k$ the normalized ones. This way, one gets discrete probability
distributions. Therefore, mean $m$ and variance $\sigma^2$ read
\begin{equation}
\begin{aligned}
&m=\sum_\ell\ell p_\ell\\
&\sigma^2=\biggl{(}\sum_\ell\ell^2p_\ell\biggr{)}-m^2\ .
\end{aligned}
\end{equation}
Using Eq.~\eqref{sigdis}, the standard deviation of the IRF can now be computed.

\medskip Taking the values from the first pass of the signals (which
refers to their structure in time, see \cite{tB97}), the standard
deviation provides already differences in dependence on the tissue
type. We tested this with some exemplary patient data. For both MR and
CT data, the values were between 1.8 and 2.6 in the case of normal
tissue and between 3.1 and 3.4 in the case of a brain tumor. This way,
getting a first indication in respect of the tissue type seems
possible. Amazingly, such information can also be extracted from CT
data, in spite of the rather short time series available.

\section{Concluding remarks}

Our brief exposition shows two things. First, a reasonable parametric
description of blood flow data, both from MR and CT, is possible and
does not suffer from standard problems of numerical deconvolution
methods.

Second, the IRF is sufficiently accessible to draw first conclusions
on the tissue type from it. In this context, we suggest the use of the
standard deviation of the IRF as a potential diagnostic parameter.

All our methods were tested on real data, which are contained in
\cite{Rost}. However, this clearly needs further investigation, and a
systematic test with a much larger data set. We hope to report on some
progress in the near future.

\subsection*{Appendix: Positivity and positive definiteness}

Let us explain the connection between these concepts, where we follow
\cite{cB75}. A complex-valued function $\varphi$ on the real line is
called {\em positive definite}, if for all $n\in{\NN}$ and for all
$n$-tupels $(x_1, \ldots, x_n)$ with $x_i\in{\RR}$, the $n\!\times\!n$
matrix $(\ \varphi(x_i-x_j)\ )_{1\leq i,j\leq n}$ is positive
Hermitian.\medskip

This definition implies the following properties:
\begin{itemize}
\item[(1)]\quad $\varphi(0)\geq 0$
\item[(2)]\quad $\varphi(-x)=\overline{\varphi(x)}$, \quad for all $x\in{\RR}$
\item[(3)]\quad $\lvert\varphi(x)\rvert\leq\varphi(0)$, \ \quad for all $x\in{\RR}$
\item[(4)]\quad $\lvert\varphi(x)-\varphi(y)\rvert^2
  \leq \ 2\ \varphi(0)\ \bigl{(}\
  \varphi(0)-\mathrm{Re}(\varphi(x-y))\ \bigr{)}$,
\end{itemize}
see \cite[Ch.~3.4]{cB75} for details. Property (4) is often called Krein's 
inequality.\medskip

A function $\varphi$ is positive definite if and only if there is a
finite positive measure $\mu$ on ${\RR}$ such that
\begin{equation}
\varphi(x)=\int_{\RR}e^{ixy}\ d\mu(y)\ .
\end{equation}
Moreover, $\varphi$ is real-valued if the measure $\mu$ is symmetric,
i.e., $\mu(A)=\mu(-A)$ for all Borel subsets of ${\RR}$. This is a
special case of Bochner's theorem, compare \cite[Thm.~3.12]{cB75}, and
$\mu$ is uniquely determined.

The special case that $\mu$ is a probability measure leads to
$\varphi(0)=1$. The function $\varphi$ is now the characteristic
function of $\mu$, and the correspondence between positive
definiteness of $\varphi$, with $\varphi(0)=1$, and $\mu$ being a
probability measure is also called the Bochner-Khintchine theorem
\cite{aS96}.

\medskip In our case, the function $\varphi$ is given and we need to
know whether the corresponding $\mu$, the IRF, is a positive signal
function. The latter is a special case of a positive measure.

If, however, any of the properties (1) -- (4) fails, we can conclude
that $\mu$ cannot be a positive measure, and hence certainly not a
positive signal.

\medskip
\subsection*{Acknowledgements}

It is a pleasure to thank E. Baake, A. Bock, P. Zeiner and N. Zint for
discussions and valuable comments on the manuscript. 
Financial support (E.R.) from Forschungsschwerpunkt Mathematisierung
(FSPM) at the University of Bielefeld is gratefully acknowledged.


\clearpage
\begin{thebibliography}{99}
\small


\bibitem{mA72}
  M.\ Abramowitz and I.\ A.\ Stegun,
  \emph{Handbook of Mathematical Functions},
  10th ed., Dover, New York (1972).

\bibitem{lA83} 
  L.\ Axel, 
  Tissue Mean Transit Time from Dynamic Computed Tomography by a 
  Simple Deconvolution Technique, 
  \emph{Investigative Radiology} {\bf 18} (1983) 94--99.

\bibitem{tB97} 
  T.\ Benner, S.\ Heiland, G.\ Erb, M.\ Forsting, and K.\ Sartor, 
 Accuracy of Gamma-Variate-Fits to Concentration-Time Curves from Dynamic
 Susceptibility-Contrast Enhanced MRI:\ Influence of Time
 Resolution, Maximal Signal Drop and Signal-To-Noise,
 \emph{Magnetic Resonance Imaging} {\bf 15} (1997) 307--317.

\bibitem{cB75}
 C.\ Berg and G.\ Forst,
 \emph{Potential Theory on Locally Compact Abelian Groups},
 Springer, Berlin (1975).

\bibitem{aB98} 
 A.\ Bock et al., 
 Quantification of the Embolization Effect in Intracranial Meningeomas 
 by $T2^*$--Weighted MR--Perfusion Measurement, 
 \emph{Klinische Neuroradiologie} {\bf 8} (1998) 30--41.

\bibitem{rB86}
 R.\ N.\ Bracewell,
 \emph{The Fourier Transform and Its Applications},
 rev.\ 2nd ed., McGraw Hill, New York (1986).


\bibitem{aC99}
 A.\ Cenic, D.\ G.\ Nabavi, R.\ A.\ Craen, A.\ W.\ Gelb and T.\ Y.\ Lee,
 Dynamic CT Measurement of Cerebral Blood Flow:\ A Validation Study,
 \emph{American Journal of Neuroradiology} {\bf20} (1999) 63--73.

\bibitem{aC00} 
 A.\ Cenic, D.\ G.\ Nabavi, R.\ A.\ Craen, A.\ W.\ Gelb and T.\ Y.\ Lee, 
 A CT Method to Measure Hemodynamics in Brain Tumors:\ Validation and 
 Application of Cerebral Blood Flow Maps, 
 \emph{American Journal of Neuroradiology} {\bf21} (2000) 462--470.

\bibitem{rD91}
 R.\ Durrett,
 \emph{Probability},
 Wadsworth, Belmont (1991).

\bibitem{aE54}
 A.\ Erd\'elyi,
 \emph{Tables of Integral Transforms},
 vol. 1, McGraw Hill, New York (1954).

\bibitem{iH55}
 I.\ I.\ Hirschman and D.\ V.\ Widder,
 \emph{The Convolution Transform},
 Princeton University Press, Princeton (1955).

\bibitem{mK87}
 M.\ Kendall and A.\ Stuart,
 \emph{Advanced Theory of Statistics},
 vol. 1, 5th ed., Charles Griffin \& Co, London (1987).

\bibitem{Rost}
 E.\ Rost,
 {\em Signalanalyse und Entfaltungsverfahren f\"ur
 Kontrastmitteluntersuchungen mittels Schnittbildverfahren},
 diploma thesis, Universit\"at Greifswald (2005).

\bibitem{aS96}
 A.\ N.\ Shiryayev,
 \emph{Probability},
 2nd ed., Springer, New York (1996).

\bibitem{dS98}
 D.\ Stark and W.\ Bradley,
 \emph{Magnetic Resonance Imaging},
 3rd ed., Elsevier Books, Oxford (1998).

\bibitem{jT} 
 J.\ K.\ Tajik, B.\ Q.\ Tran and E.\ A.\ Hoffmann,
 New Technique to Quantitate Regional Pulmonary
 Microvascular Transit Times from Dynamic X-ray CT Images,
 in:\ Proceedings of the Pacific Medical Technology Symposium (1998), 
 R.\ Nelson, A.\ Gelish and S.\ K.\ Mun (eds.), 
 available online\footnote{available at:
 http://csdl.computer.org/comp/\linebreak[0]procee\-dings/pacmedtek/1998/8667/00/8667toc.htm.}.

\bibitem{rW00} 
 R.\ Wirestam, L.\ Andersson, L.\ \O{}stergaard, M.\ Bolling, J.-P.\ Aunola, 
 A.\ Lindgren, B.\ Geijer, S.\ Holt\aa{}s and F.\ St\aa{}hlberg, 
 Assessment of Regional Cerebral Blood Flow by Dynamic Susceptibility 
 Contrast MRI Using Different Deconvolution Techniques, 
 \emph{Magnetic Resonance in Medicine} {\bf 43} (2000) 691--700.

\end{thebibliography}
\end{document}